\documentclass[10pt, conference]{IEEEtran}
\IEEEoverridecommandlockouts
\usepackage{cite}
\usepackage{amsmath,amssymb,amsfonts}
\usepackage{algorithmic}
\usepackage{graphicx}
\usepackage{textcomp}
\usepackage{hyperref}
\usepackage{xcolor}
\usepackage{multirow, multicol}
\usepackage{booktabs} 
\def\BibTeX{{\rm B\kern-.05em{\sc i\kern-.025em b}\kern-.08em
    T\kern-.1667em\lower.7ex\hbox{E}\kern-.125emX}}
    
\begin{document}

\title{Transformation of audio embeddings into interpretable, concept-based representations}


\author{\IEEEauthorblockN{Alice Zhang\textsuperscript{*}\thanks{*Work completed during internship at Dolby Laboratories}}
\IEEEauthorblockA{\textit{Dept. of Electrical and Computer Engineering} \\
\textit{Univ. of Texas at Austin}\\
Austin, USA \\
alice.zhang@austin.utexas.edu}
\and
\IEEEauthorblockN{Edison Thomaz}
\IEEEauthorblockA{\textit{Dept. of Electrical and Computer Engineering} \\
\textit{Univ. of Texas at Austin}\\
Austin, USA \\
ethomaz@utexas.edu}
\and
\IEEEauthorblockN{Lie Lu}
\IEEEauthorblockA{\textit{Dolby Laboratories} \\
San Francisco, USA \\
llu@dolby.com}
}

\maketitle

\begin{abstract}
Advancements in audio neural networks have established state-of-the-art results on downstream audio tasks. However, the black-box structure of these models makes it difficult to interpret the information encoded in their internal audio representations. In this work, we explore the semantic interpretability of audio embeddings extracted from these neural networks by leveraging CLAP, a contrastive learning model that brings audio and text into a shared embedding space. We implement a post-hoc method to transform CLAP embeddings into concept-based, sparse representations with semantic interpretability. Qualitative and quantitative evaluations show that the concept-based representations outperform or match the performance of original audio embeddings on downstream tasks while providing interpretability. Additionally, we demonstrate that fine-tuning the concept-based representations can further improve their performance on downstream tasks. Lastly, we publish three audio-specific vocabularies for concept-based interpretability of audio embeddings.


\end{abstract}


\section{Introduction}
\label{sec:intro}
Neural networks for audio recognition and classification have improved significantly in recent years with the development of models, such as the convolution neural network(CNN)-based family of Pretrained Audio Neural Networks (PANNs) \cite{kong2020panns} and the transformer-based Hierarchical Token-Semantic Audio Transformer (HTS-AT) \cite{htsat-ke2022}. These models not only established state-of-the-art (SOTA) results in audio tasks, such as sound event detection and text-audio retrieval, but have also formed a basis for subsequent multimodal models. For instance, the audio language model Pengi \cite{deshmukh2023pengi} uses the HTS-AT as its audio encoder to realize downstream tasks ranging from audio captioning to music analysis. Despite the advantages of these neural networks, their black-box structure makes it difficult to understand their inner audio representations. In this work, we aim to establish semantic interpretability of audio embeddings extracted from these models. To achieve this, we draw on text embeddings, which encode semantic information and enable alignment between audio representations and linguistic meaning. Since audio and text embeddings exist in different spaces (spectrogram space versus latent space of a deep network), we leverage contrastive learning to address this gap.

Contrastive learning learns a representation such that similar pairs of data points are closer together in a shared embedding space while dissimilar pairs are farther apart. It has become increasingly popular for multimodal representation learning, galvanized by models such as CLIP \cite{radford2021clip} for image-text pairs and CLAP \cite{CLAP2022,laionclap2023} for audio-text pairs. CLAP models, which leverage either CNN14 from the PANNs family or HTS-AT as their audio encoder, have been trained to provide the advantage of a joint multimodal latent space that yields semantically-rich representations of audio data. This shared latent space enables efficiency and scalability on high-performing downstream tasks, such as zero-shot classification and information retrieval \cite{radford2021clip, laionclap2023,CLAP2023}. Despite the semantic information encoded in the audio embeddings, the embeddings within the latent space are not easily interpretable to humans. In this work, we aim to explore the question: \textit{how can we understand the semantics of the audio data encoded in CLAP embeddings?}

     
\begin{figure*}
    \centering
    \includegraphics[width=1.0\textwidth]{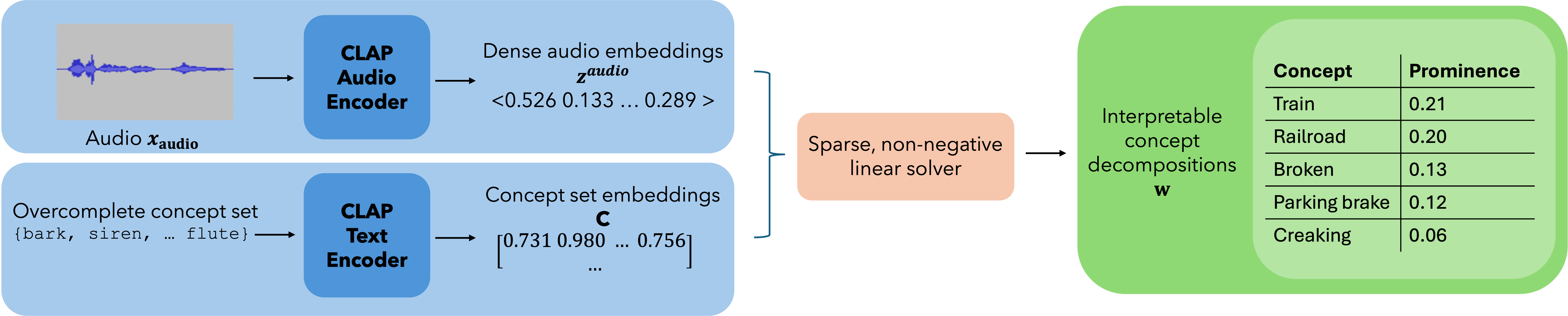}
    \caption{A diagram of our concept decomposition system illustrates how dense CLAP embeddings (z) are transformed into concept-based representation (w) by solving for a sparse, non-negative linear decomposition over a concept vocabulary (C).}
    \label{fig:method}
    \vspace{-10pt}
\end{figure*} 

As machine learning becomes increasingly deployed to real-world systems, explainability is an integral component of its responsible use, if not a legal requirement altogether \cite{goodman2017european}. While there have been increased efforts for interpretability in the computer vision and image domain \cite{gandelsman2024interpreting, bhalla2024interpretingclipsparselinear, materzynska2022disentangling}, there has been significantly less work in interpretability and explainability for audio machine learning models. In this work, we aim to utilize the multimodal nature of CLAP embeddings to transform dense CLAP embeddings into a sparse, human-interpretable representation. Specifically, our contributions increase the interpretability of audio embeddings:
     \begin{enumerate}
         \item We explore a \textit{post-hoc} method to transform original CLAP embeddings into concept-based representations comprising a sparse combination of interpretable, semantic concepts in a computationally efficient manner in section \ref{sec:method}. Additionally, we present an initial exploration in fine-tuning the concept-based representations to further improve the performance of concept-based representations on downstream tasks.
         \item We perform an extensive set of experiments demonstrating our concept-based representations improve upon the downstream performance of original CLAP embeddings while providing added interpretability in section \ref{sec:experiments}. Furthermore, we investigate factors, such as concept set construction methods and number of concepts, that further the performance of our concept-based representations on downstream tasks.
         \item We create three audio-specific vocabularies for concept-based interpretability of audio embeddings and make them publicly available \href{https://osf.io/3cgsu/?view_only=ecbf92d4b10a48a38441323fc275a97f}{here}\footnote{\url{https://osf.io/3cgsu/?view_only=ecbf92d4b10a48a38441323fc275a97f}}. 
     \end{enumerate}

\section{Related Work}

\subsection{Audio Interpretability}
Prior works have utilized visualization techniques to highlight input spectrogram features that contribute to a model's decision \cite{won2019interpretable, Becker2018InterpretingAE}. However, visualization of audio as time-frequency images provides limited interpretability to a general user \cite{Liberman1968WhyAS}. Additional work operates on audio as spectrogram-like 2D images and leverages image-based approaches for interpretability such as feature perturbation, perturbing the input and observing the changes in the output\cite{fucci2024spesspectrogramperturbationexplainable}. 

Other works have built upon Local Interpretable Model-agnostic Explanations (LIME) \cite{ribeiro2016whyitrustyou}, a feature-attribution method that treats machine learning models as a black box and explains a model's prediction by observing outputs of the black box in response to a large number of inputs. SoundLIME \cite{Mishra2017LocalIM} localizes the time or time-frequency region in an input spectrogram that contribute most to a model's decision, and audioLIME \cite{haunschmid2020audiolimelistenableexplanationsusing} creates listenable interpretations through source separation. While audioLIME generates more interpretable explanations than SoundLIME, audioLIME depends on a source separation system that works with a limited number of predefined audio sources and is therefore not easily scalable to diverse audio sounds or datasets. Parekh et al. \cite{parekh2022listen} also create listenable interpretations using non-negative matrix factorization to learn a spectral pattern dictionary and then decompose an audio signal into its constituent spectral patterns.

In this work, we implement a \textit{post-hoc} explainability approach to explain a trained model rather than build an explainable model by-design\cite{ghorbani2019automatic}. To address the limited scalability of existing methods, our method is \textit{concept-based} and task-agnostic, with high-level human-friendly concepts that can more easily scale to various audio sounds and sources. Our concept-based method provides the advantage of directly understanding the semantic content within audio embeddings whereas methods such as perturbation of counterpart text embeddings are indirect and assume that the text embeddings well capture nuances of the audio signal.

\subsection{Concept Bottleneck Models}
Concept bottleneck models (CBMs) are a family of interpretable models that map input features onto a set of interpretable concepts and then express their prediction as a linear combination of the concepts \cite{koh2020cbm, oikarinen2023labelfreeconceptbottleneckmodels, yuksekgonul2023posthoc}. However, these models require expert-labeled concept datasets for training. While recent works on CBMs have leveraged querying large language models (LLMs) to obtain concept datasets, these concept datasets are subject to the biases of LLMs \cite{oikarinen2023labelfreeconceptbottleneckmodels, chattopadhyay2024information, panousis2023sparse}. Furthermore, CBMs often do not match the performance of unrestricted neural networks \cite{yuksekgonul2023posthoc}. In our work, we create a large-scale and overcomplete concept dictionary that does not require specific domain knowledge. Additionally, we demonstrate that our concept-based representations of audio embeddings match or improve upon the performance of corresponding dense audio embeddings. Lastly, prior work has focused primarily on CBMs for interpretability of image tasks with little work on audio tasks \cite{koh2020cbm, oikarinen2023clipdissect, yuksekgonul2023posthoc, oikarinen2023labelfreeconceptbottleneckmodels}.

\section{Method}
\label{sec:method}

Our method, which is inspired by the SpLiCE \cite{bhalla2024interpretingclipsparselinear} method for interpreting CLIP (image) embeddings, is illustrated in Figure \ref{fig:method}. The inputs are an audio waveform and a vocabulary of natural language concepts. The output is a sparse vector in which each dimension represents a concept and most dimensions are zero. By removing the zero elements, the output can be further simplified as a compact set of concepts that semantically represent the input audio. 

\subsection{Concept Vocabulary Construction}
\label{vocab}
Prior work in interpretability established three desired properties of \textit{concept-based} explanations of machine learning models: \textit{meaningfulness} - providing standalone semantic definition, \textit{coherency} - instances of a concept should be similar to each other and different from instances of other concepts, and \textit{importance} - the concept is necessary for the true prediction of samples in a class \cite{ghorbani2019automatic}. Here, we define concept as a semantic unit expressed by an English word, and we use combinations of semantic concepts expressed as natural language to meet these desiderata.

For the baseline vocabulary, we want a concept set that is not task specific and instead covers a large lexicon of sounds. Therefore, we scrape the audio tags of the FSD50K dataset\cite{fonseca2022fsd50k}, which is a human-labeled dataset with over 50,000 audio samples spanning a variety of sound sources including animals, humans and machines. Out of an initial total of 20,793 unique, human-provided audio tags, we sort the tags by frequency, remove expletives, and finally select the 2,000 most frequent English audio tags. We also build two alternative concept vocabularies from this baseline vocabulary: 1) a \textit{pruned} vocabulary and 2) a \textit{clustered} vocabulary. 

For the \textit{pruned} vocabulary, we first consider the 10,000 most frequent audio tags in the FSD50K dataset. Then, we filter the concepts to remove mis-spelled English words, single letter audio tags, and numeric audio tags. We also identify synonym concepts and concepts sharing the same root word and keep only the most frequent audio tag representing the concept. For instance, the concepts ``cough”, ``coughs,” and ``coughing” appear in the 10,000 most frequent audio tags of FSD50K. We keep only the concept ``cough” to represent all the three concepts since it appears most frequently.  After aggregating all synonyms and root words and their frequency counts, we keep the 2,000 most frequent concepts for the pruned vocabulary. 

For the \textit{clustered} vocabulary, we also first consider the 10,000 most frequent audio tags in the FSD50K dataset. Then, for all 10,000 audio tags, we cluster the 1,024-dimension text embedding for each tag obtained via the CLAP text encoder into 2,000 clusters via k-means clustering. For each cluster, we select the concept with the text embedding closest to the cluster's centroid to serve as the cluster's representative concept. 

We choose to manually create our concept set over querying LLMs, as recent work has shown that concept sets generated via LLMs are reliant on the domain knowledge and subject to the biases of LLMs. As a result, LLMs may fail to generate concepts important to certain classes \cite{oikarinen2023labelfreeconceptbottleneckmodels, chattopadhyay2024information, panousis2023sparse}. Therefore, we choose to create an over-complete vocabulary set so that the concept set is task-agnostic. To maintain interpretability with the over-complete set, we enforce sparsity in the concept decomposition. 

\begin{figure}
    \centering
    \includegraphics[width=0.9\linewidth]{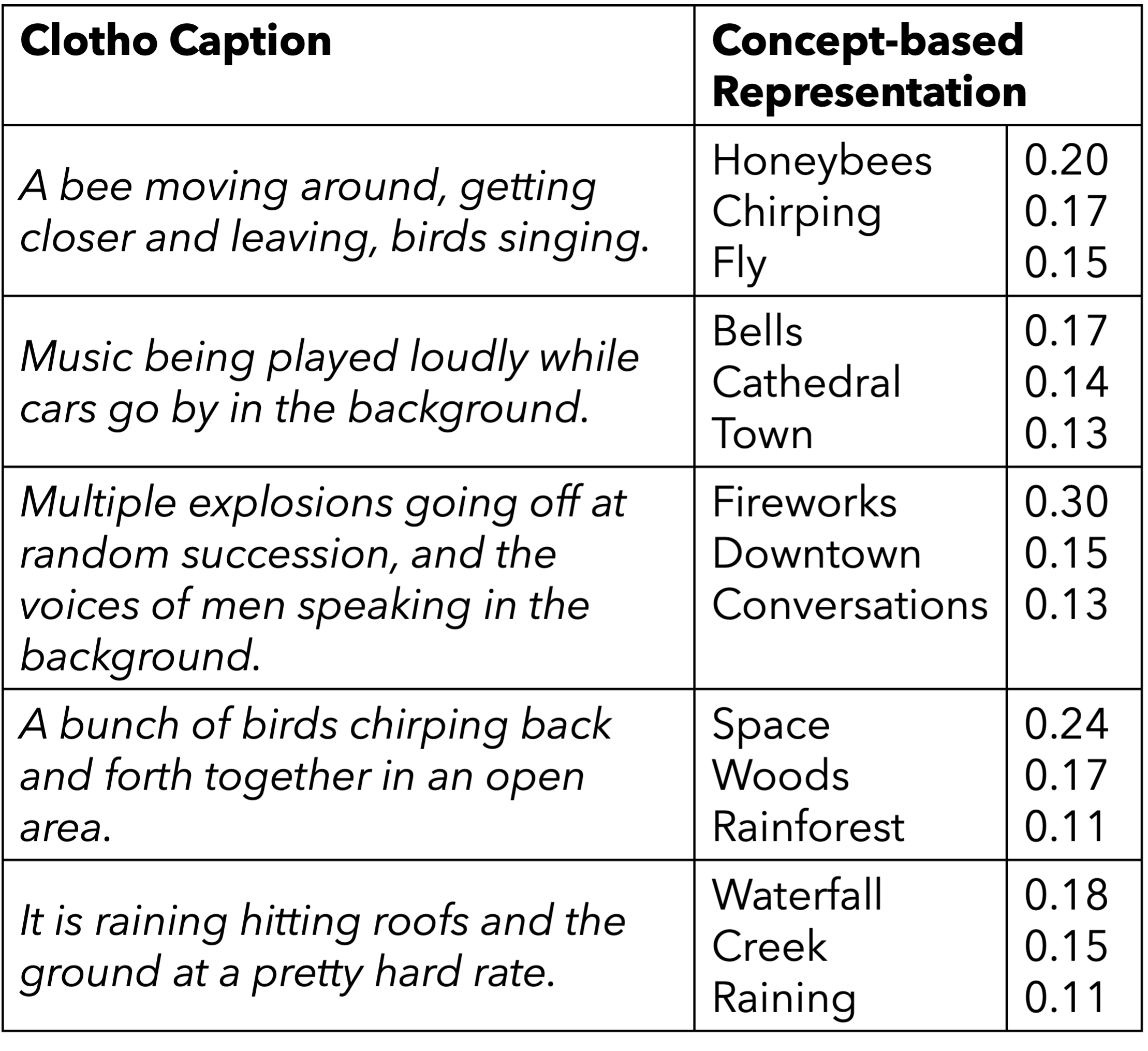}
    \caption{Example audios from Clotho with their captions and corresponding concept representation (concept, prominence value) of audio signals. We show the top-3 concepts but the audio embedding decompositions have a total of 35-45 concepts.}
    \label{fig:examples}
    \vspace{-10pt}
\end{figure}
\subsection{Sparse Linear Embedding Decomposition}
We use the general method of sparse linear embedding decomposition from SpLiCE \cite{bhalla2024interpretingclipsparselinear} and apply it to audio embeddings. Let \(\mathbf{x}^{audio}\) and \(\mathbf{x}^{concept}\) be a raw audio sample and a concept respectively.
Given a CLAP audio encoder $f:\mathbb{R}^{d_a} \rightarrow \mathbb{R}^{d}$  and text encoder $g:\mathbb{R}^{d_t} \rightarrow \mathbb{R}^{d}$, we define CLAP representations in $\mathbb{R}^d$ as $\textbf{z}^{audio}=f(\textbf{x}^{audio})$ and $\textbf{z}^{concept}=g(\textbf{x}^{concept})$. 

Then, the goal is to approximate $\textbf{z}^{audio} \approx \textbf{Cw}^*$ where $\mathbf{C}=\{\mathbf{z_1}^{concept}, ..., \mathbf{z_c}^{concept}\}\in \mathbb{R}^{d*c}$ is a fixed vocabulary with $c$ concepts, and $\textbf{w}^* \in \mathbb{R}^c$ is the concept-based decomposition.

We can find a sparse solution vector by minimizing the L0 norm of the vector $\textbf{w}^*$ such that the cosine similarity between the original audio embedding and the reconstructed embedding through concept-based representation $\textbf{Cw}^*$ is greater than $1- \epsilon$ for some small $\epsilon$. This is defined formally as:	
\[{w}^* = \min_{w\in\mathbb{R}^c} \|\mathbf{w}\|_0  s.t. \langle \mathbf{z}^{audio}, \mathbf{Cw} \rangle \geq(1-\epsilon) \]for some small $\epsilon$.
We relax the L0 constraint as is standard practice due to the non-convexity of the L0 norm. We reformulate the objective as minimizing the convex L1 norm using Lasso regression, defined as: 
\[{w}^* = \min_{w\in\mathbb{R}^c} \| \mathbf{Cw} - \mathbf{z}^{audio}\|_2^2 +\lambda\|\mathbf{w}\|_1 \]
In its implementation, we use sklearn's Lasso solver with a non-negativity flag to enforce non-negativity. Therefore, the solution to this equation is a sparse, non-negative vector where non-zero values correspond to the prominence of concepts present in the original audio sample or audio embedding. The hyperparameter $\lambda$ (L1 penalty) determines the number of non-zero concepts in the concept-based representation. While other sparse, linear solvers exist such as orthogonal matching pursuit, the guarantee of non-negative weights in the solution vector aids in the understanding of concepts present in the input audio. Through subjective human evaluations, prior works have indicated that concepts with non-negative weights are easier to understand and more meaningful than concepts with negative weights \cite{mutahar2022conceptbasedexplanationsusingnonnegative, zhang2020improving}.  

\subsection{Fine-Tuned Sparse Embedding Decomposition}
We can further fine-tune the original CLAP audio embedding to a specific downstream task prior to embedding decomposition. We project the original audio embedding with a single linear layer $H \in  \mathbb{R}^{d*d}$ where $H$ is initialized as a random $ d\times d$ matrix. In other words, we use $H$ to map $\textbf{z}^{audio}$ onto the text embeddings extracted from the text prompt corresponding to a downstream task, and then solve for the vector of weights that best approximates the projected audio embedding $H\textbf{z}^{audio}$. The solution we seek is then:
\[{w}^* = \min_{w\in\mathbb{R}^c} \| \mathbf{Cw} - H\mathbf{z}^{audio}\|_2^2 +\lambda\|\mathbf{w}\|_1 \]
Section \ref{fine-tuned-eval} provides additional implementation details. 


\begin{figure}
    \centering
    \includegraphics[width=0.65\linewidth]{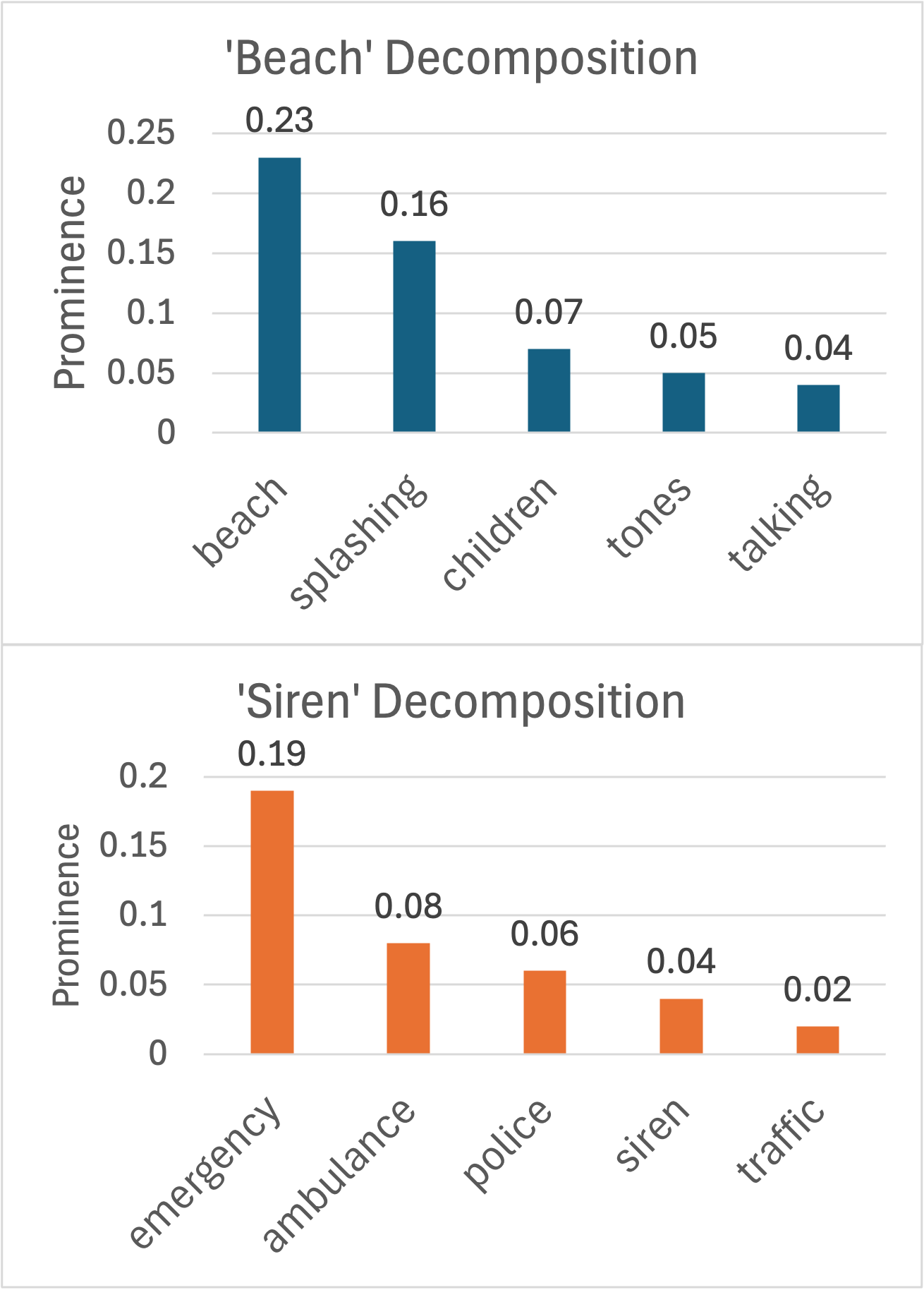}
    \caption{Distribution of top-5 concepts across two audio classes.}
    \label{fig:hist}
    \vspace{-10pt}

\end{figure}

\section{Experiments}
\label{sec:experiments}
There are two existing CLAP models, one trained by LAION on 630K audio-text pairs \cite{laionclap2023} and one trained by Microsoft on 4.6M audio-text pairs \cite{CLAP2023}. Since Microsoft's CLAP model was trained on a larger dataset, we focus our evaluations on Microsoft's CLAP model. We  note, however, that our proposed method works with either LAION's or Microsoft's CLAP model. We evaluate our method qualitatively and quantitatively on 7 datasets (Urbansound8K, DCASE2017 Task 4 Subtask A, ESC-50, AudioSet, Vocalsound, TUT2017, and Clotho) from 5 different domains (sound event classification, vocal sound classification, acoustic scene classification, audio-text and text-audio retrieval) as downstream tasks. 

\begin{table*}[htbp]
    \centering
    \scriptsize
    \caption{Performance of concept-based representations on zero-shot classification tasks. We benchmark against the SOTA zero-shot results in literature and the original CLAP model. Evaluation metrics are F1-score for DCASE17 (imbalanced dataset), mAP for AudioSet, and accuracy for all other datasets. Confidence intervals obtained from bootstrap sampling each dataset's evaluation set.}
    \begin{tabular}{c|c|c|c|c|c|c} \hline 
         &  \multicolumn{4}{c|}{Sound Event Classification}&  Vocal Sound Classification& Acoustic Scene Classification\\ \hline 
         Model&  Urbansound8K \cite{Salamon:UrbanSound:ACMMM:14} & DCASE2017 Task 4 \cite{Mesaros2019sound} & ESC-50 \cite{piczak2015dataset} & AudioSet \cite{audioset} & Vocalsound \cite{Gong2022vocalsound} & TUT2017 \cite{Mesaros2019sound} \\ 
         \hline
         1. Benchmark&  0.806 \cite{mei2024wavcaps} &  0.3 \cite{CLAP2022} & \textbf{0.948} \cite{mei2024wavcaps}& \textbf{0.277} \cite{zhu2024languagebindextendingvideolanguagepretraining}& \textbf{0.849} \cite{ma2024investigatingemergentaudioclassification}& 0.296 \cite{CLAP2022} \\ 
         2. CLAP \cite{CLAP2023} & 0.823 & 0.466 & 0.939 & 0.268 & 0.8 & 0.538\\ 
         3. Concept-Based Rep. (ours) & \textbf{0.828 $\pm$ 0.006}& \textbf{0.47 $\pm$ 0.021}& 0.937 $\pm$0.011& 0.265 $\pm$ 0.007&  0.821 $\pm$ 0.011& \textbf{0.556 $\pm$ 0.027}\\
    \end{tabular}
    \label{tab:classification}
    \vspace{-10pt}
\end{table*}

\begin{table*}[htbp]    
    \centering
    \scriptsize
    \caption{Performance of concept-based representations on zero-shot information retrieval tasks. We benchmark against SOTA systems that use the HTS-AT architecture as the audio encoder for fair comparison to the CLAP model evaluated in this work, which also uses the HTS-AT architecture as the audio encoder.}
    \begin{tabular}{c|c c|c c} \hline 
         &  \multicolumn{2}{ c |}{Audio-Text Retrieval}&  \multicolumn{2}{c}{Text-Audio Retrieval}\\ \hline 
         Model&  R@1&  mAP@10&  R@1& mAP@10\\ \hline 
         1. Benchmark& 0.234 \cite{mei2024wavcaps} & 0.138 \cite{laionclap2023} & \textbf{0.195} \cite{mei2024wavcaps} & 0.204\cite{laionclap2023}\\ 
         2. CLAP \cite{CLAP2023} & 0.229 & \textbf{0.155} & 0.157 & 0.257\\ 
         3. Concept-Based Rep. (ours)& \textbf{0.240} & 0.151 & 0.162 & \textbf{0.261}\\
    \end{tabular}
    \label{tab:retrieval}
    \vspace{-10pt}
    
\end{table*}

\subsection{Qualitative Evaluation}
We qualitatively evaluate the audio concept decompositions by the semantic meaning of the extracted concepts representing the input audio sample. The qualitative evaluations were conducted with an L1 penalty of 0.15 which yielded 35-45 non-zero concepts in the concept-based representation with a vocabulary size of 2,000. An example of the top-5 non-zero concepts extracted from an audio sample of a train is illustrated in Figure \ref{fig:method}. In Figure \ref{fig:examples}, we provide 5 concept-representations of audio samples from the Clotho audio-captioning dataset \cite{drossos2019clothoaudiocaptioningdataset} with their corresponding captions. We find that the concepts describe the audio content as indicated by the original captions and can provide semantic explanations of CLAP embeddings. Notably, this qualitative evaluation demonstrates that our concept-based representations are able to capture semantic concepts corresponding to multiple audio sources within an audio embedding with the sparsity constraint. Additionally, the decomposition method works with audio of varying lengths, from 1 to 15 seconds. 


We further extend our decomposition method to entire sound classes or datasets to gain a better understanding of a collection of audio samples without needing to listen to each sample individually. We perform concept decomposition on each audio sample within a target class or dataset and average the prominence values of each concept across all audio samples to better understand the semantic distribution of the datasets. For instance, in the ``beach" class in TUT2017, the top concepts besides ``beach" are ``splashing" and ``children," suggesting that the audio samples were collected from recreational beaches rather than wildlife beaches. We visualize the distribution of top-5 concepts for this class in addition to the ``siren" class in Urbansound8K in Figure \ref{fig:hist}. This decomposition at the class level shows at a glance the siren sounds in the dataset are primarily from ambulance or police vehicles. Despite the availability of other concepts in the vocabulary that can explain siren sources, such as weather warnings or sports game celebrations, these concepts are omitted from the decomposition and highlight the utility of class-level decomposition.

\subsection{Quantitative Zero-Shot Evaluation}
\label{quantitative evaluation}
We evaluate the proposed method on audio classification and information retrieval to show the effectiveness of our concept-based representations on downstream tasks and verify that the additional interpretability does not compromise downstream tasks. For initial quantitative evaluations, we fix the L1 penalty at 0.05 resulting in 90-100 non-zero concepts and use the baseline vocabulary with a size of 2,000 concepts.

\textbf{Zero-Shot Classification}
To evaluate the performance of concept-based representations on zero-shot classification, we calculate the cosine similarity between the concept-based representations, which can be viewed as a concept-based reconstruction of the original audio embedding, and text embeddings encoding the class label prompt. The logits are transformed into probability distributions by applying a softmax for multiclass classification. For fair comparison to the performance of original, dense CLAP audio embeddings, we follow the CLAP evaluation setup and use the prompt, ``This is a sound of [class label]." 

We summarize the results of zero-shot classification for sound event detection, vocal sound detection and acoustic scene classification in Table \ref{tab:classification} and compare the results to those of SOTA zero-shot methods and original CLAP embeddings. As shown, the performance of concept-based representations consisting of 90-100 non-zero concepts in zero-shot classification outperform that of the original CLAP embeddings for Urbansound8K, DCASE2017, Vocalsound, and TUT2017. With the exception of Vocalsound, these three datasets also surpass benchmark zero-shot results for their respective datasets. These results demonstrate that concept-based representations with added interpretability can also improve the performance of downstream classification tasks. On datasets with a larger number of sound event classes such as ESC-50 and AudioSet, which have 50 and 527 sound event labels respectively, the concept-based representations very closely approach the performance of the original CLAP embeddings. These results show that, even on large-vocabulary datasets, adding interpretability to audio emebddings does not take away from their performance. 

 \begin{table*}[!htbp]
    \centering
    \caption{Performance of fine-tuned concept decomposition, comparing with state-of-the-art \textit{supervised} methods and fine-tuned CLAP embedding. Evaluation metric is accuracy.}
    \scriptsize
    \begin{tabular}{c|c|c|c|c} \hline 
         & \multicolumn{2}{c|}{Sound Event Classification}&  Vocal Sound Classification& Acoustic Scene Classification\\ \hline 
         Model&  Urbansound8K \cite{Salamon:UrbanSound:ACMMM:14} & ESC-50 \cite{piczak2015dataset}  & Vocalsound \cite{Gong2022vocalsound} & TUT2017 \cite{Mesaros2019sound} \\ 
         \hline
         1. SoTA Supervised &  \textbf{0.9007} \cite{guzhov2021audioclipextendingclipimage} & \textbf{0.991} \cite{omnivec2}& \textbf{0.929} \cite{Qwen-Audio}& \textbf{0.649} \cite{Qwen-Audio}\\ 
         2. Fine-Tuned CLAP Embeddings (No decomposition) & 0.897 $\pm$ 0.021& 0.972 $\pm$ 0.012& 0.865 $\pm$ 0.01& 0.615 $\pm$ 0.019\\
         3. Fine-Tuned Concept-Based Rep. & 0.9 $\pm$ 0.023& 0.969 $\pm$ 0.012& 0.855 $\pm$ 0.011& 0.647 $\pm$ 0.021\\
    \end{tabular}
    \label{tab:supervised-classification-results}
    \vspace{-10pt}
 \end{table*} 
 
 \textbf{Zero-Shot Information Retrieval} Similar to zero-shot classification, we compute the cosine similarity between the concept-based representation and the text embeddings encoding the text query to determine the best audio-text pair. We evaluate on Clotho and compare our results to SOTA zero-shot methods and original CLAP embeddings. The results are summarized in Table \ref {tab:retrieval}. We observe mixed performance with our concept-based representations, with our method improving the R@1 and mAP@10 of audio-text and text-audio retrieval respectively compared to the benchmarks. 

\begin{figure}[h!]
    \centering
    \includegraphics[width=1.0\linewidth, scale=1.0]{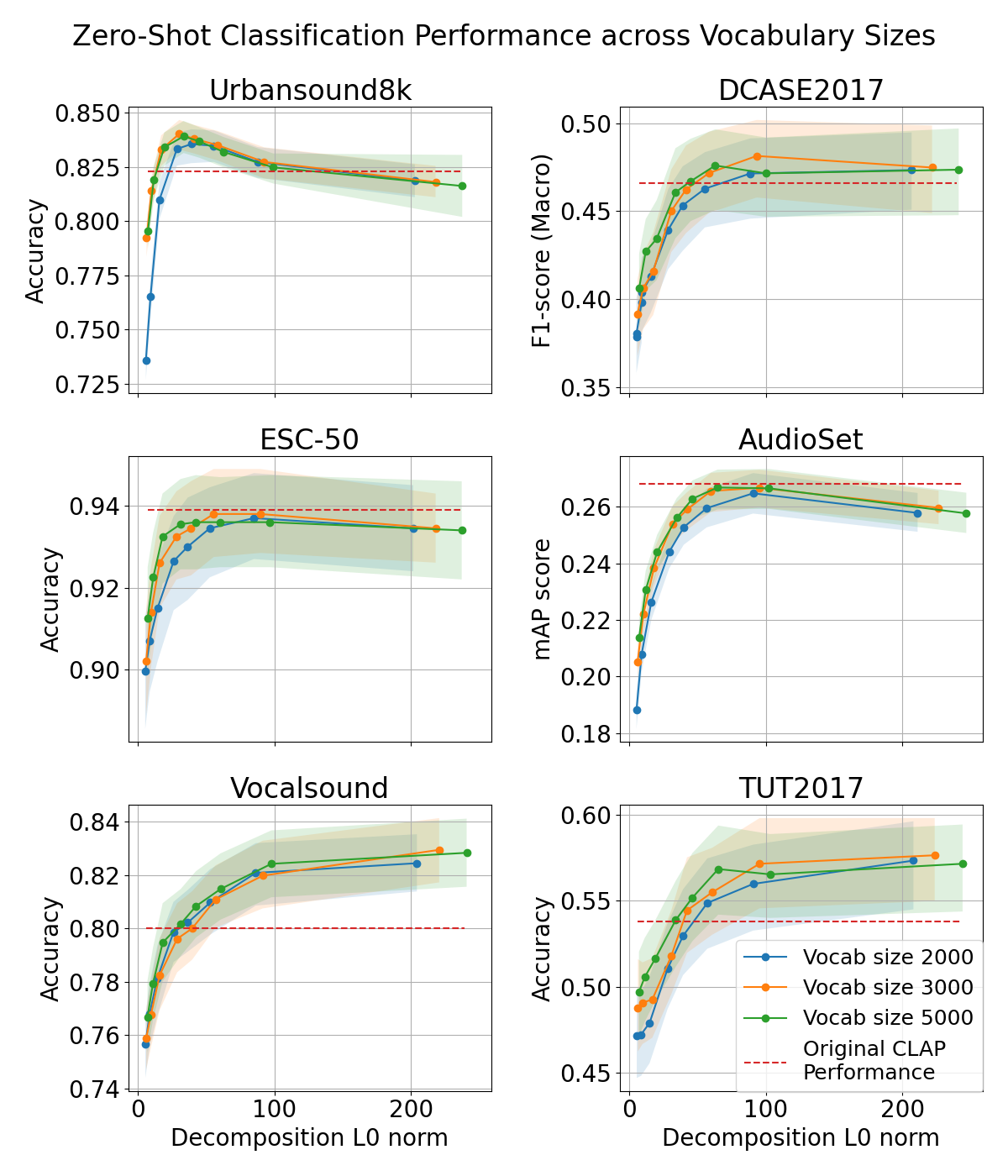}
    \caption{Zero-shot classification on multiple datasets as the L1 penalty varies from 0.01 to 0.50, resulting in solutions with L0 norms between $\sim$5-200 and as the vocabulary size varies from 2,000 to 5,000 concepts.}
    \label{fig:sparsity-classification}
\vspace{-10pt}    
\end{figure}

\subsection{Fine-Tuned Decomposition Evaluation}
\label{fine-tuned-eval}
We evaluate the effectiveness of the fine-tuned audio embedding decomposition on four datasets (Urbansound8K, ESC-50, Vocalsound, and TUT2017), with a focus on examining how downstream task performance of datasets with fewer training samples can benefit from this fine-tuning. We train the linear projection layer to maximize the cosine similarity between the original, dense CLAP embedding extracted from the CLAP audio encoder and the text embedding extracted from the CLAP text encoder that encodes the prompt ``This is a sound of [class label]." We train and evaluate the projection layer using the defined development/evaluation splits or folds corresponding to each dataset. After training the projection layer, we use the layer to transform the original CLAP audio embeddings in the evaluation set and then perform embedding decomposition to obtain concept-based representations of the projected CLAP embeddings. Similar to the zero-shot setups, we determine the cosine similarity between the concept-based representations of the projected audio embeddings and text embeddings of the class prompts to obtain a prediction for each audio sample and calculate the final performance metric.




 \begin{figure}[h!]
    \centering
    \includegraphics[width=1.0\linewidth, scale=1.0]{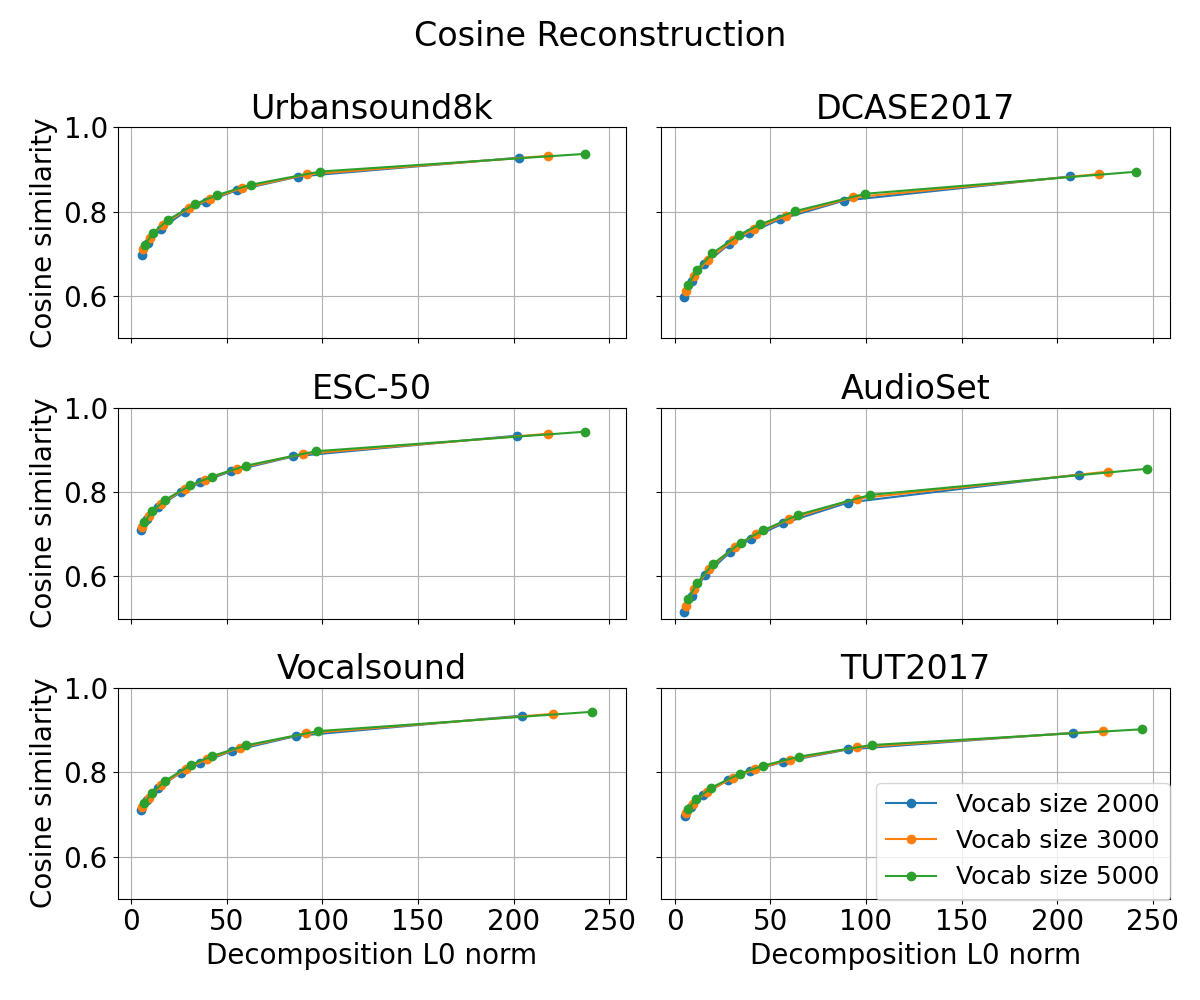}
    \caption{Cosine similarity between the concept-based representation and original CLAP embedding.}
    \label{fig:sparsity-cosine}
\vspace{-10pt}    
\end{figure}

\begin{figure}[h!]
    \centering
    \includegraphics[width=1\linewidth, scale=1]{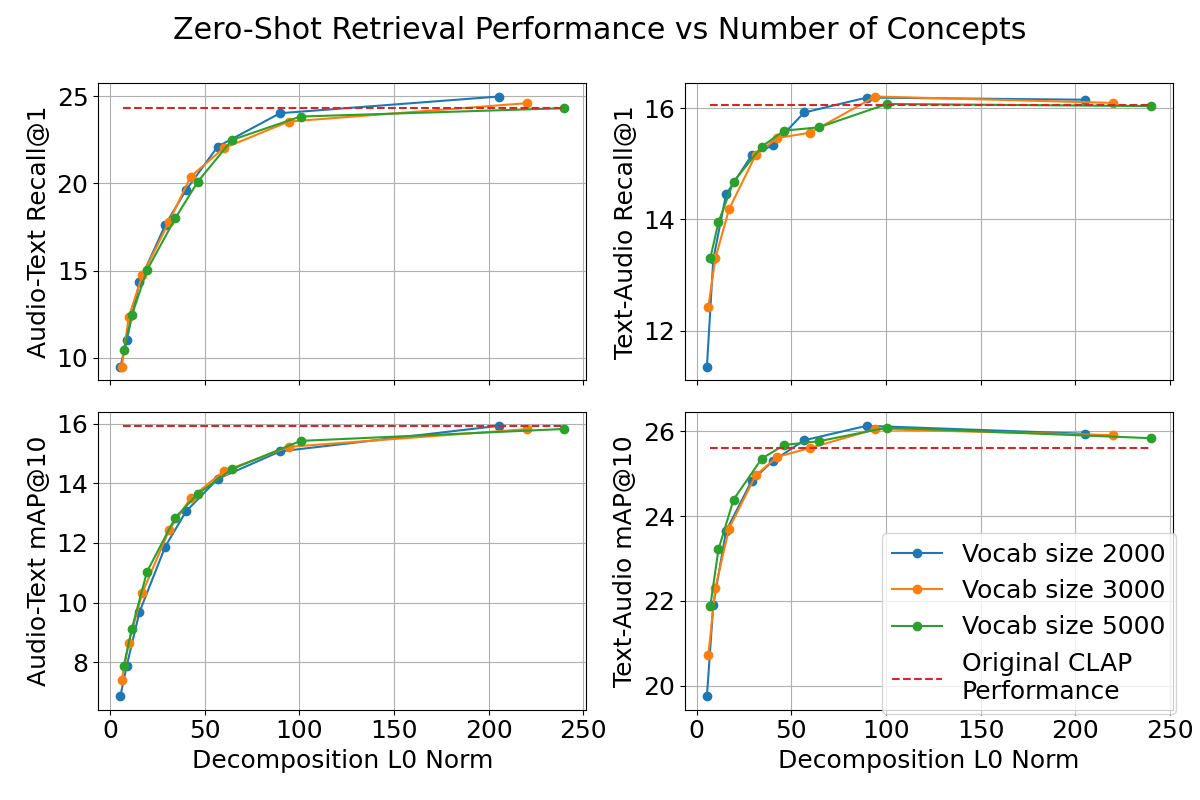}
    \caption{Zero-shot information retrieval on the Clotho dataset as the L1 penalty varies from 0.01 to 0.50, resulting in solutions with L0 norms between $\sim$5-200 and as the vocabulary size varies from 2,000 to 5,000 concepts.}
    \label{fig:sparsity-retrieval}
\vspace{-10pt}    
\end{figure}

\begin{table*}[!htbp]
    \centering
    \caption{Concept decomposition across three vocabularies for an audio embedding from Clotho dataset with original caption ``A person is pouring something metal into a dish."}
    \begin{tabular}{c|ccc} \hline
    & Baseline vocabulary & Pruned vocabulary & Clustered vocabulary \\
    \hline
    \multirow{3}*{Top-3 concepts, prominence} & rainstick, 0.170 & coffee, 0.130 & coin-spinning, 0.170  \\
        & gold, 0.112  & bongo, 0.100 & rattling, 0.151  \\
        & rattling, 0.096 & rattle, 0.098 & corn, 0.129   \\
    \hline
    Cosine similarity  & 0.856 & 0.871 & 0.857\\
    \end{tabular}
    \label{tab:vocab-comp}
\vspace{-10pt}    
 \end{table*}


The comparison results are shown in Table \ref{tab:supervised-classification-results}. We observe that the projected CLAP embedding followed by concept decomposition (row 3) improves upon the performance of downstream tasks on average, compared to the fined-tuned CLAP embedding (without concept decomposition, row 1); it also has significant improvement compared to the results without the projection layer in Table \ref{tab:classification} (row 3). With a single linear projection layer, it seems to match the SoTA \textit{supervised} audio classification methods for Urbansound8K and TUT2017 datasets at accuracies 0.9 and 0.646 respectively. Our results also show an accuracy gap in the VocalSound dataset, indicating room for improvement in the use of supervision for concept-decomposition, such as the use of more complex models or different loss functions. 
 
\subsection{Sparsity-Performance Tradeoffs}
We investigate the relationship between the number of non-zero concepts used in building concept-based representations and their downstream zero-shot classification performance by sweeping the L1 penalty between [0.01, 0.5]. As shown in Figure \ref{fig:sparsity-classification}, beginning with $\sim$40 concepts and above (correlating to L1 penalties less than 0.15), performance of zero-shot classification using concept-based representation for all datasets except ESC-50 and AudioSet surpasses that of original CLAP embeddings. Using only $\sim$20-40 non-zero concepts in our representations, we can achieve similar performance as the dense CLAP embeddings, displaying a significant reduction in memory while maintaining performance compared to original CLAP embeddings. 

Interestingly, we observe that performance does not always improve with a greater number of non-zero concepts, as seen with Urbansound8K, ESC-50, and AudioSet. For ESC-50 and AudioSet, we hypothesize this is partly due to their large-label nature. With granular classes that span multiple sound categories, it is possible that additional concepts do not contribute meaningfully to the audio representation for downstream classification. However, further research is required to better understand this trend. 

In Figure \ref{fig:sparsity-cosine}, we also investigate the impact of the L1 penalty on the cosine similarity between the concept-based representation and the CLAP audio embedding. As expected, the cosine similarity increases as the number of non-zero concepts representing the audio embedding increases.

For the downstream retrieval tasks, the recall rate of the concept-based decomposition with an L1 penalty of 0.05 ($\sim$100 non-zero concepts) surpasses that of the original CLAP embedding on average (Table \ref {tab:retrieval}). As the L1 penalty increases to 0.1 ($\sim$60 non-zero concepts) and above, retrieval performance decreases significantly as shown in Figure \ref{fig:sparsity-retrieval}. 
\label{sparsity-tradeoff}

Figures \ref{fig:sparsity-classification}, \ref{fig:sparsity-cosine}, and \ref{fig:sparsity-retrieval} also show results for additional vocabulary sizes which will be discussed in the following section.

\subsection{Effects of Concept Set Construction and Size}
As mentioned in section \ref{vocab}, we construct a baseline vocabulary of 2,000 concepts from audio tags in the FSD50K dataset. Here, we consider the effect of larger vocabulary sizes. Additionally, we build two additional \textit{pruned} and \textit{clustered} vocabularies of 2,000 concepts by filtering and clustering audio tags in the FSD50K dataset.

\textbf{Effects of Concept Set Size}
While we initially use the 2,000 most frequent audio tags in FSD50K, we now examine using a larger subset of the most frequent audio tags in FSD50K. We therefore consider the 3,000 and 5,000 most frequent audio tags in FSD50K without any pruning or clustering. We find that the size of the vocabulary used to decompose the audio embedding has a minimal impact on downstream task performance (Figures \ref{fig:sparsity-classification} and \ref{fig:sparsity-retrieval}). Furthermore, Figure \ref{fig:sparsity-cosine} indicates that the cosine similarity between the concept-based representation and the original CLAP audio embedding does not vary significantly as the size of the vocabulary changes. 

\textbf{Effects of concept set construction} We qualitatively and quantitatively examine the performance of the three different vocabulary sets on audio embedding decomposition. To reiterate, the vocabularies differ in their method of construction but all three vocabularies have 2,000 concepts. Qualitatively, we show the top-3 concepts extracted by each vocabulary set from an audio file containing sounds of objects pouring into a dish from the Clotho dataset in Table \ref{tab:vocab-comp}. The three concept sets capture similar semantic ideas and reconstruct well the original CLAP embedding (cosine similarity $>$ 0.85). Notably, all three concept sets identify variations of the concept ``rattle" as a top concept. The pruned vocabulary identifies ``rattle" unlike the baseline and clustered vocabularies, which identify ``rattling", because ``rattling" was removed in favor of ``rattle" in the pruned vocabulary. 

\begin{figure}[h!]
    \centering
    \includegraphics[width=1.0\linewidth, scale=1]{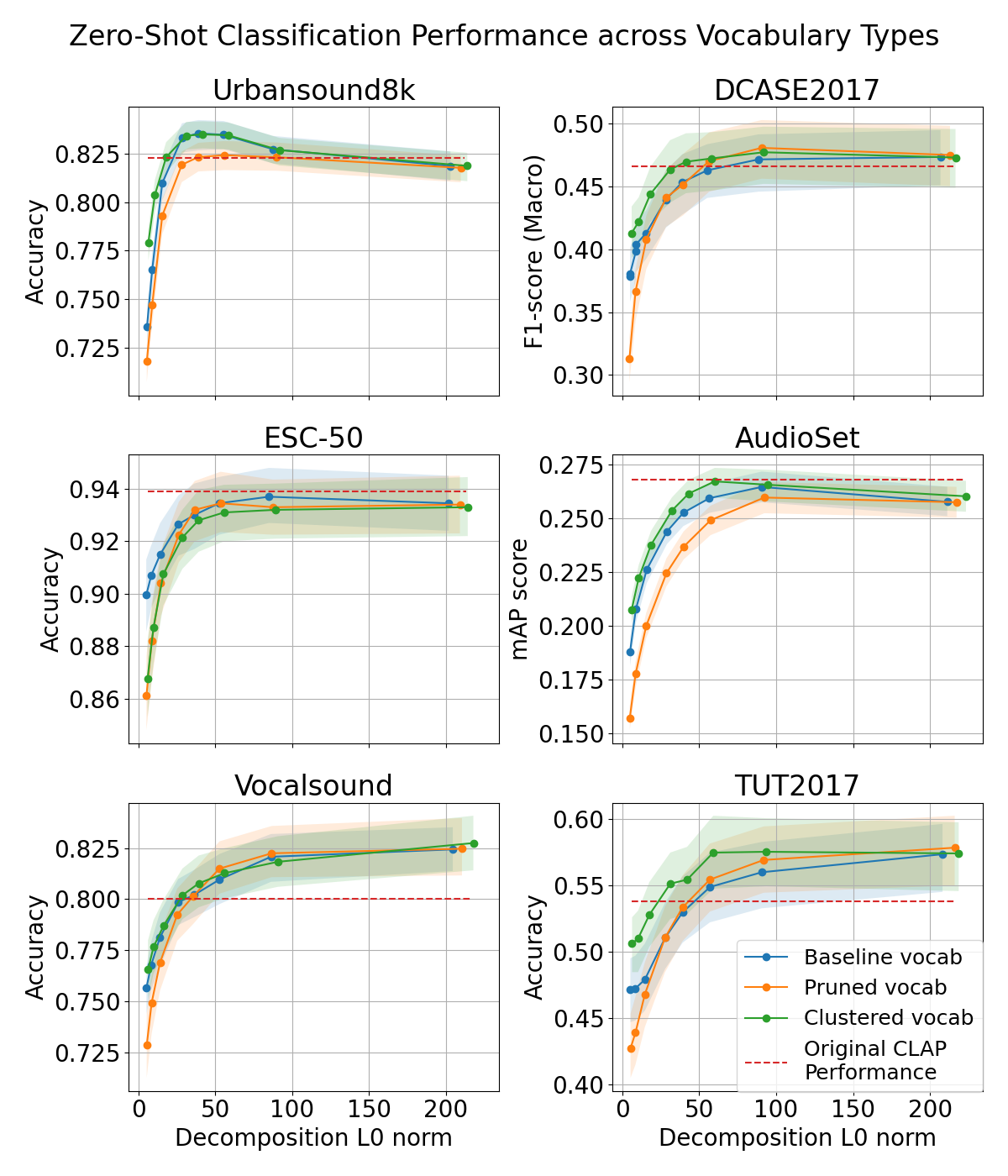}
    \caption{Zero-shot classification as the L1 penalty varies from 0.01 to 0.50 using a constant vocabulary size of 2,000 concepts across three concept sets.}
    \label{fig:vocab-type-classification}
\vspace{-10pt}    
\end{figure}

\begin{figure}[h!]
    \centering
    \includegraphics[width=1\linewidth, scale=1]{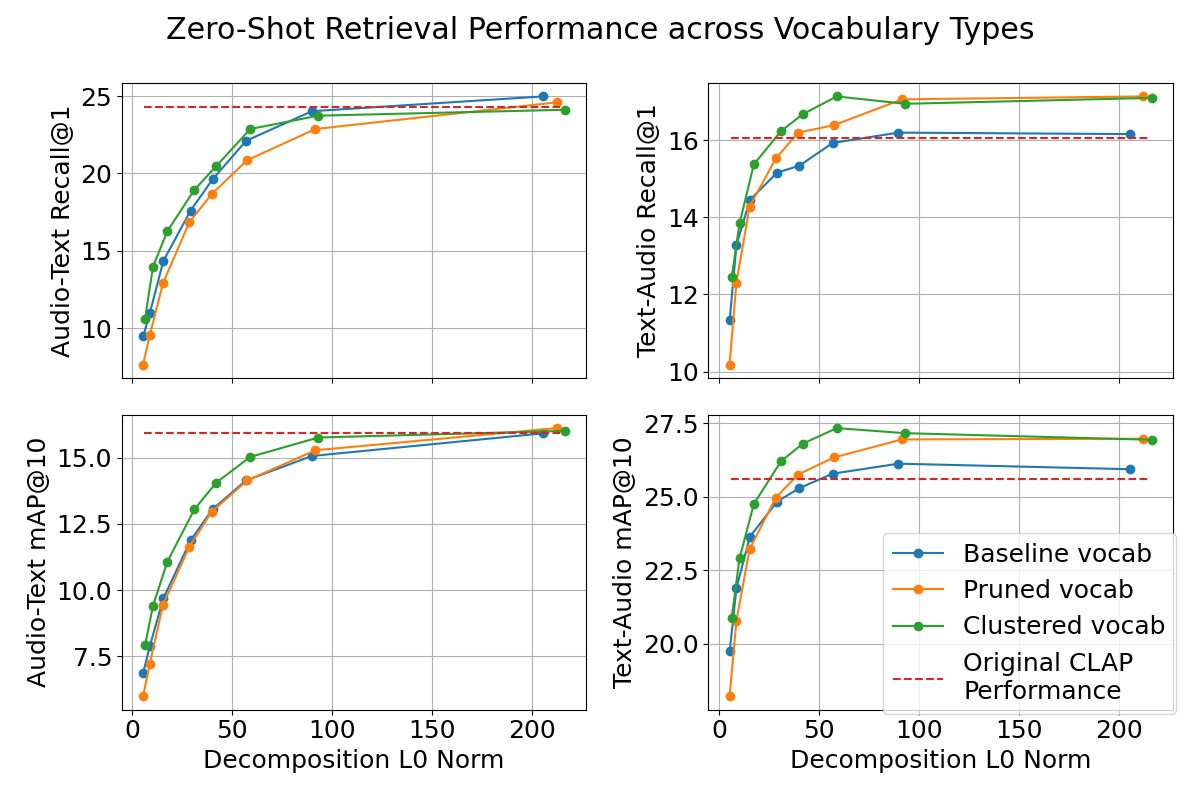}
   \caption{Zero-shot information retrieval on the Clotho dataset as the L1 penalty varies from 0.01 to 0.50 using a constant vocabulary size of 2,000 concepts across three concept sets.}
    \label{fig:vocab-type-retrieval}
\vspace{-10pt}    
\end{figure}

Quantitatively, we examine the performance of zero-shot classification and retrieval using the three concept sets with the L1 penalty sweeping across [0.01, 0.5] as in section \ref{sparsity-tradeoff}.  Figure \ref{fig:vocab-type-classification} shows the baseline and clustered vocabulary generally outperform the pruned vocabulary for zero-shot classification. In contrast, Figure \ref{fig:vocab-type-retrieval} shows the clustered and pruned vocabulary outperforming the baseline vocabulary for text-to-audio retrieval. However, there is no significant performance difference between the vocabularies for audio-text retrieval. We hypothesize that the improvement in using the pruned or clustered vocabularies over the baseline vocabulary for the text-to-audio retrieval task is due to the pruned and clustered vocabularies spanning a larger concept space to better represent the original CLAP audio embeddings. Despite all vocabularies having 2,000 concepts, the pruned and clustered vocabularies have a set of more unique concepts, since their redundant concepts have been removed. Consequently, CLAP audio embeddings represented by richer semantic concepts allow for a given text query to better find the audio file with the content specified by the text query.

\section{Conclusion}
\label{sec:conclusion}
In this paper, we introduce a method to transform original CLAP embeddings into concept-based representations for increased interpretability of audio embeddings, which has been understudied in comparison to image and text embeddings. Our approach removes the need to collect labeled data for predefined concepts, which is time consuming and labor intensive, and a limitation of existing audio interpretability methods. We demonstrate that the concept-based representations improve or match the performance of original CLAP embeddings on downstream classification and retrieval tasks. Our concept-based representations enable future work in concept-based audio editing or generation.

\bibliographystyle{IEEEtran}
\bibliography{refs}

\begin{thebibliography}{10}
\providecommand{\url}[1]{#1}
\csname url@samestyle\endcsname
\providecommand{\newblock}{\relax}
\providecommand{\bibinfo}[2]{#2}
\providecommand{\BIBentrySTDinterwordspacing}{\spaceskip=0pt\relax}
\providecommand{\BIBentryALTinterwordstretchfactor}{4}
\providecommand{\BIBentryALTinterwordspacing}{\spaceskip=\fontdimen2\font plus
\BIBentryALTinterwordstretchfactor\fontdimen3\font minus \fontdimen4\font\relax}
\providecommand{\BIBforeignlanguage}[2]{{%
\expandafter\ifx\csname l@#1\endcsname\relax
\typeout{** WARNING: IEEEtran.bst: No hyphenation pattern has been}%
\typeout{** loaded for the language `#1'. Using the pattern for}%
\typeout{** the default language instead.}%
\else
\language=\csname l@#1\endcsname
\fi
#2}}
\providecommand{\BIBdecl}{\relax}
\BIBdecl

\bibitem{kong2020panns}
Q.~Kong, Y.~Cao, T.~Iqbal, Y.~Wang, W.~Wang, and M.~D. Plumbley, ``{PANN}s: Large-scale pretrained audio neural networks for audio pattern recognition,'' \emph{IEEE/ACM TASLPRO}, vol.~28, pp. 2880--2894, 2020.

\bibitem{htsat-ke2022}
K.~Chen, X.~Du, B.~Zhu, Z.~Ma, T.~Berg-Kirkpatrick, and S.~Dubnov, ``{HTS-AT}: A hierarchical token-semantic audio transformer for sound classification and detection,'' in \emph{{ICASSP} 2022}, 2022.

\bibitem{deshmukh2023pengi}
S.~Deshmukh, B.~Elizalde, R.~Singh, and H.~Wang, ``Pengi: An audio language model for audio tasks,'' in \emph{NeurIPS}, vol.~36, 2023, pp. 18\,090--18\,108.

\bibitem{radford2021clip}
A.~Radford, J.~W. Kim, C.~Hallacy, A.~Ramesh, G.~Goh, S.~Agarwal, G.~Sastry, A.~Askell, P.~Mishkin, J.~Clark \emph{et~al.}, ``Learning transferable visual models from natural language supervision,'' in \emph{ICML}.\hskip 1em plus 0.5em minus 0.4em\relax PmLR, 2021, pp. 8748--8763.

\bibitem{CLAP2022}
B.~Elizalde, S.~Deshmukh, M.~Al~Ismail, and H.~Wang, ``{CLAP}: learning audio concepts from natural language supervision,'' in \emph{ICASSP 2023}.\hskip 1em plus 0.5em minus 0.4em\relax IEEE, 2023, pp. 1--5.

\bibitem{laionclap2023}
Y.~Wu*, K.~Chen*, T.~Zhang*, Y.~Hui*, T.~Berg-Kirkpatrick, and S.~Dubnov, ``Large-scale contrastive language-audio pretraining with feature fusion and keyword-to-caption augmentation,'' in \emph{ICASSP}, 2023.

\bibitem{CLAP2023}
B.~Elizalde, S.~Deshmukh, and H.~Wang, ``Natural language supervision for general-purpose audio representations,'' in \emph{ICASSP 2024}, 2024, pp. 336--340.

\bibitem{goodman2017european}
B.~Goodman and S.~Flaxman, ``European union regulations on algorithmic decision making and a “right to explanation”,'' \emph{AI Magazine}, vol.~38, no.~3, p. 50–57, Sep. 2017.

\bibitem{gandelsman2024interpreting}
Y.~Gandelsman, A.~A. Efros, and J.~Steinhardt, ``Interpreting {CLIP}'s image representation via text-based decomposition,'' in \emph{ICLR 2024}, 2024.

\bibitem{bhalla2024interpretingclipsparselinear}
U.~Bhalla, A.~Oesterling, S.~Srinivas, F.~Calmon, and H.~Lakkaraju, ``Interpreting {CLIP} with sparse linear concept embeddings ({S}p{LICE}),'' \emph{NeurIPS 2024}, vol.~37, pp. 84\,298--84\,328, 2024.

\bibitem{materzynska2022disentangling}
J.~Materzyńska, A.~Torralba, and D.~Bau, ``Disentangling visual and written concepts in {CLIP},'' in \emph{CVPR 2022}, 2022, pp. 16\,389--16\,398.

\bibitem{won2019interpretable}
M.~Won, S.~Chun, and X.~Serra, ``Toward interpretable music tagging with self-attention,'' \emph{arXiv}, 2019.

\bibitem{Becker2018InterpretingAE}
S.~Becker, M.~Ackermann, S.~Lapuschkin, K.~M{\"u}ller, and W.~Samek, ``Interpreting and explaining deep neural networks for classification of audio signals,'' \emph{ArXiv}, 2018.

\bibitem{Liberman1968WhyAS}
A.~M. Liberman, F.~S. Cooper, D.~P. Shankweiler, and M.~Studdert-Kennedy, ``Why are speech spectrograms hard to read?'' \emph{American annals of the deaf}, vol. 113 2, pp. 127--33, 1968.

\bibitem{fucci2024spesspectrogramperturbationexplainable}
D.~Fucci, M.~Gaido, B.~Savoldi, M.~Negri, M.~Cettolo, and L.~Bentivogli, ``{SPES}: Spectrogram perturbation for explainable speech-to-text generation,'' \emph{arXiv}, 2024.

\bibitem{ribeiro2016whyitrustyou}
M.~T. Ribeiro, S.~Singh, and C.~Guestrin, ``"{W}hy should {I} trust you?": Explaining the predictions of any classifier,'' in \emph{Proceedings of the 22nd ACM SIGKDD International Conference on Knowledge Discovery and Data Mining}.\hskip 1em plus 0.5em minus 0.4em\relax ACM, 2016, p. 1135–1144.

\bibitem{Mishra2017LocalIM}
S.~Mishra, B.~L. Sturm, and S.~Dixon, ``Local interpretable model-agnostic explanations for music content analysis,'' in \emph{ISMIR}, 2017.

\bibitem{haunschmid2020audiolimelistenableexplanationsusing}
V.~Haunschmid, E.~Manilow, and G.~Widmer, ``{audio{LIME}: Listenable Explanations Using Source Separation},'' 13th International Workshop on Machine Learning and Music, 2020.

\bibitem{parekh2022listen}
J.~Parekh, S.~Parekh, P.~Mozharovskyi, F.~d\textquotesingle Alch\'{e}-Buc, and G.~Richard, ``Listen to interpret: Post-hoc interpretability for audio networks with {NMF},'' in \emph{NeurIPS}, vol.~35, 2022, pp. 35\,270--35\,283.

\bibitem{ghorbani2019automatic}
A.~Ghorbani, J.~Wexler, J.~Y. Zou, and B.~Kim, ``Towards automatic concept-based explanations,'' in \emph{NeurIPS 2019}, 2019, pp. 9273--9282.

\bibitem{koh2020cbm}
P.~W. Koh, T.~Nguyen, Y.~S. Tang, S.~Mussmann, E.~Pierson, B.~Kim, and P.~Liang, ``Concept bottleneck models,'' in \emph{ICML 2020}, vol. 119.\hskip 1em plus 0.5em minus 0.4em\relax PMLR, 2020, pp. 5338--5348.

\bibitem{oikarinen2023labelfreeconceptbottleneckmodels}
T.~Oikarinen, S.~Das, L.~M. Nguyen, and T.-W. Weng, ``Label-free concept bottleneck models,'' in \emph{ICLR 2023}, 2023.

\bibitem{yuksekgonul2023posthoc}
M.~Yuksekgonul, M.~Wang, and J.~Zou, ``Post-hoc concept bottleneck models,'' in \emph{ICLR 2023}, 2023.

\bibitem{chattopadhyay2024information}
A.~Chattopadhyay, R.~Pilgrim, and R.~Vidal, ``Information maximization perspective of orthogonal matching pursuit with applications to explainable {AI},'' in \emph{NeurIPS 2023}, 2023.

\bibitem{panousis2023sparse}
K.~P. Panousis, D.~Ienco, and D.~Marcos, ``{ Sparse Linear Concept Discovery Models },'' in \emph{ICCV 2013}, 2023, pp. 2759--2763.

\bibitem{oikarinen2023clipdissect}
T.~Oikarinen and T.-W. Weng, ``{CLIP}-{D}issect: Automatic description of neuron representations in deep vision networks,'' in \emph{ICLR 2023}, 2023.

\bibitem{fonseca2022fsd50k}
E.~Fonseca, X.~Favory, J.~Pons, F.~Font, and X.~Serra, ``{FSD50K}: An open dataset of human-labeled sound events,'' \emph{IEEE/ACM Trans. Audio, Speech and Lang. Proc.}, vol.~30, p. 829–852, Dec. 2021.

\bibitem{mutahar2022conceptbasedexplanationsusingnonnegative}
G.~Mutahar and T.~Miller, ``Concept-based explanations using non-negative concept activation vectors and decision tree for {CNN} models,'' \emph{arXiv}, 2022.

\bibitem{zhang2020improving}
R.~Zhang, P.~Madumal, T.~Miller, K.~A. Ehinger, and B.~I. Rubinstein, ``Invertible concept-based explanations for {CNN} models with non-negative concept activation vectors,'' in \emph{AAAI 2021}, vol.~35, no.~13, 2021, pp. 11\,682--11\,690.

\bibitem{Salamon:UrbanSound:ACMMM:14}
J.~Salamon, C.~Jacoby, and J.~P. Bello, ``A dataset and taxonomy for urban sound research,'' in \emph{ACM Multimedia}, Orlando, FL, USA, Nov. 2014, pp. 1041--1044.

\bibitem{Mesaros2019sound}
A.~Mesaros, A.~Diment, B.~Elizalde, T.~Heittola, E.~Vincent, B.~Raj, and T.~Virtanen, ``Sound event detection in the {DCASE} 2017 challenge,'' \emph{IEEE/ACM TASLPRO}, 2019, in press.

\bibitem{piczak2015dataset}
K.~J. Piczak, ``{ESC}: {Dataset} for {Environmental Sound Classification},'' in \emph{ACM Multimedia 2015}.\hskip 1em plus 0.5em minus 0.4em\relax {ACM Press}, pp. 1015--1018.

\bibitem{audioset}
J.~F. Gemmeke, D.~P.~W. Ellis, D.~Freedman, A.~Jansen, W.~Lawrence, R.~C. Moore, M.~Plakal, and M.~Ritter, ``Audio {S}et: An ontology and human-labeled dataset for audio events,'' in \emph{ICASSP 2017}, 2017, pp. 776--780.

\bibitem{Gong2022vocalsound}
Y.~Gong, J.~Yu, and J.~Glass, ``Vocalsound: A dataset for improving human vocal sounds recognition,'' in \emph{ICASSP 2022}.\hskip 1em plus 0.5em minus 0.4em\relax IEEE, 2022.

\bibitem{mei2024wavcaps}
X.~Mei, C.~Meng, H.~Liu, Q.~Kong, T.~Ko, C.~Zhao, M.~D. Plumbley, Y.~Zou, and W.~Wang, ``{W}av{C}aps: A {C}hat{GPT}-assisted weakly-labelled audio captioning dataset for audio-language multimodal research,'' \emph{IEEE/ACM TASLPRO}, vol.~32, p. 3339–3354, 2024.

\bibitem{zhu2024languagebindextendingvideolanguagepretraining}
B.~Zhu, B.~Lin, M.~Ning, Y.~Yan, J.~Cui, H.~Wang, Y.~Pang, W.~Jiang, J.~Zhang, Z.~Li, W.~Zhang, Z.~Li, W.~Liu, and L.~Yuan, ``Language{B}ind: Extending video-language pretraining to n-modality by language-based semantic alignment,'' \emph{arXiv}, 2024.

\bibitem{ma2024investigatingemergentaudioclassification}
R.~Ma, A.~Liusie, M.~Gales, and K.~Knill, ``Investigating the emergent audio classification ability of {ASR} foundation models,'' in \emph{2024 NAACL: Human Language Technologies (Volume 1: Long Papers)}.\hskip 1em plus 0.5em minus 0.4em\relax Mexico City, Mexico: ACL, Jun. 2024, pp. 4746--4760.

\bibitem{drossos2019clothoaudiocaptioningdataset}
K.~Drossos, S.~Lipping, and T.~Virtanen, ``Clotho: an audio captioning dataset,'' in \emph{ICASSP 2020}, 2020, pp. 736--740.

\bibitem{guzhov2021audioclipextendingclipimage}
A.~Guzhov, F.~Raue, J.~Hees, and A.~Dengel, ``Audio{CLIP}: Extending {CLIP} to image, text and audio,'' in \emph{ICASSP 2022}, 2022, pp. 976--980.

\bibitem{omnivec2}
S.~Srivastava and G.~Sharma, ``Omni{V}ec2 - a novel transformer based network for large scale multimodal and multitask learning,'' in \emph{2024 CVPR}, 2024, pp. 27\,402--27\,414.

\bibitem{Qwen-Audio}
Y.~Chu, J.~Xu, X.~Zhou, Q.~Yang, S.~Zhang, Z.~Yan, C.~Zhou, and J.~Zhou, ``Qwen-audio: Advancing universal audio understanding via unified large-scale audio-language models,'' \emph{arXiv preprint arXiv:2311.07919}, 2023.

\end{thebibliography}

\end{document}